\documentclass[longauth,traditabstract]{aa} 
\usepackage{graphicx}
\usepackage{txfonts}
\begin{document}
   \title{Discovery of very high energy gamma-ray emission from the blazar 1ES 1727+502 with the MAGIC Telescopes}

%
\author{
 J.~Aleksi\'c\inst{1} \and
 L.~A.~Antonelli\inst{2} \and
 P.~Antoranz\inst{3} \and
 M.~Asensio\inst{4} \and
 M.~Backes\inst{5} \and
 U.~Barres de Almeida\inst{6} \and
 J.~A.~Barrio\inst{4} \and
 J.~Becerra Gonz\'alez\inst{7} \and
 W.~Bednarek\inst{8} \and
 K.~Berger\inst{7,}\inst{9}
% ...
}
\institute { IFAE, Edifici Cn., Campus UAB, E-08193 Bellaterra, Spain
 \and INAF National Institute for Astrophysics, I-00136 Rome, Italy
 \and Universit\`a  di Siena, and INFN Pisa, I-53100 Siena, Italy
 \and Universidad Complutense, E-28040 Madrid, Spain
 \and Technische Universit\"at Dortmund, D-44221 Dortmund, Germany
 \and Max-Planck-Institut f\"ur Physik, D-80805 M\"unchen, Germany
 \and Inst. de Astrof\'{\i}sica de Canarias, E-38200 La Laguna, Tenerife, Spain
 \and University of \L\'od\'z, PL-90236 Lodz, Poland
 \and Depto. de Astrof\'{\i}sica, Universidad de La Laguna, E-38206 La Laguna, Spain
% ...
}

\date{Received ... ; accepted ...  }

 \abstract{Motivated by the Costamante \& Ghisellini (2002) predictions we investigated if the blazar 1ES 1727+502 ($z=0.055)$ is emitting very high energy (VHE, E$>$100\,GeV) $\gamma$ rays. 
  We observed the BL Lac object 1ES 1727+502 in stereoscopic mode with the two MAGIC telescopes during 14 nights between May 6th and June 10th 2011, for a total effective observing time 
of 12.6 hours. For the study of the multiwavelength spectral energy distribution (SED) we use simultaneous optical \textit{R}-band data from the KVA telescope, archival UV/optical and X-ray 
observations by instruments UVOT and XRT on board of the \textit{Swift} satellite and high energy (HE, 0.1\,GeV - 100\,GeV) $\gamma$--ray data from the \emph{Fermi}-LAT instrument.
 We detect, for the first time, VHE $\gamma$--ray emission from 1ES 1727+502 at a statistical significance of 5.5\,$\sigma$. The integral flux above 150\,GeV is estimated to be $(2.1\pm0.4)\%$ of the Crab Nebula flux and the de-absorbed VHE spectrum has a photon index of $(2.7\pm0.5)$. No significant short-term variability was found in any of the wavebands presented 
 here. We model the SED using a one-zone synchrotron self-Compton model obtaining parameters typical for this class of sources.} 


	\titlerunning{1ES 1727+502}

   \maketitle
%

\section{Introduction}

Since the detection of the first extragalactic VHE $\gamma$--ray source, Mrk 421 in 1992 by the Whipple Observatory (Punch et al. 1992), 
the extragalactic VHE sky turned out to be densely populated. Currently, around 50 extragalactic sources\footnote{http://tevcat.uchicago.edu/} 
are known, most of them blazars, i.e. Active Galactic Nuclei (AGN) with a relativistic jet pointed towards the Earth. Blazars can be further divided into 
BL Lacertae objects (BL Lacs) and Flat Spectrum Radio Quasars (FSRQs). The former class constitutes the vast majority of blazars detected so far in the VHE $\gamma$--ray regime. 
Their spectral energy distributions (SEDs) are characterized by two broad peaks, located in the radio - IR - optical - UV - X-ray regime and the HE - VHE $\gamma$--ray bands respectively. 
BL Lacs are further divided into high frequency peaked BL Lacs (HBL) and low frequency peaked BL Lacs (LBL, Padovani \& Giommi 1995). Their emission is generally believed to 
be caused by a population of relativistic electrons, trapped in a region with magnetic field, that emit synchrotron photons, forming the low-energy peak. Those photons are then 
up-scattered to higher energies by the same population of electrons, through the inverse Compton process to form the second bump (SSC, Synchrotron Self Compton scenario).

Imaging atmospheric Cherenkov telescopes carry on pointed observations in search for extragalactic sources and do not perform scans of the entire sky due to their limited field of view ($\sim$ 3.5$^\circ$). 
The selection of promising candidates for VHE emission is thus of fundamental importance. 
The BL Lac object 1ES 1727+502 (discuss in this paper) is the latest in a long list of MAGIC discoveries of objects selected from X-ray catalogues (e.g. for 1ES 1727+502 
Costamante \& Ghisellini 2002, but for other sources also Donato et al. 2001). Among those are 1ES 1218+30.4 (Albert et al. 2006a), PG 1553+113 (Albert et al. 2007a), 
1ES 1741+196 (Berger et al. 2011) and 1ES 0033+595 (Mariotti et al. 2011). Also many of the sources, whose discoveries have been triggered by an optical high state (Mrk180, 
Albert et al. 2006a; 1ES 1011+496, Albert et al. 2007b; B3 2247+381, Aleksi\'c et al. 2012a; 1ES 1215+303, Aleksi\'c et al. 2012b) are listed in the above mentioned catalogues.

The BL Lac 1ES 1727+502 ($z = 0.055$, de Vaucouleurs et al. 1991) was observed with the Whipple 10\,m $\gamma$--ray telescope, in March-April 1995 and April-May 1996, for a total of 4.6 hours, but no signal from this source
was detected. Upper limits above 300\,GeV were reported for both data sets at the level of $1.08\times 10^{-11}\,\mathrm{erg}\,\mathrm{cm}^{-2}\,\mathrm{s}^{-1}$  (8.6$\%$ Crab), and 
$1.58\times 10^{-11}\,\mathrm{erg}\,\mathrm{cm}^{-2}\,\mathrm{s}^{-1}$  (15$\%$ Crab), respectively (Horan et al. 2004). It has also been a target studied with the single 
telescope MAGIC-I (Albert et al. 2008a), before starting stereoscopic observations with two MAGIC telescopes (Aleksi\'c et al. 2012c). 
It was observed between May 2006 and May 2007 for $\sim$ 6.1 hours, with zenith angles from 21$^\circ$ to 36$^\circ$. 
An upper limit on the integral flux of $3.6\times 10^{-11}\,\mathrm{cm}^{-2}\,\mathrm{s}^{-1}$ above 140\,GeV  (11.8\% of the Crab Nebula flux above 140\,GeV) was calculated. 
These observations were merged with the ones from 20 other pre-selected blazars observed between 2004 and 2009, and analysed with a stacking method (Aleksi\'c et al. 2011). 
The combined dataset with 394.1 hours exposure time resulted in a detection of VHE $\gamma$ rays with a statistical significance of 4.9\,$\sigma$, thus indicating that at least some of those blazars are VHE $\gamma$--ray emitters. 
In June 2010, a high optical flux of 1ES 1727+502 triggered target of opportunity observations with the MAGIC telescopes. Unfortunately, data were unusable due to adverse atmospheric 
conditions.

The hard spectrum in the HE band (spectral index 2.0 in the \textit{Fermi}-LAT first source catalogue, Abdo et al. 2010), combined with the better sensitivity achieved by the MAGIC telescopes with respect to the one of 2006 and 2007 observations, motivated renewed MAGIC observations in 2011, which are described in the following sections.
In the second \textit{Fermi}-LAT catalog (Nolan et al. 2012)  the object 1ES 1727+502 (2FGL J1728.2+5015) confirmed a hard spectrum with spectral index of 1.8.

\section{MAGIC observations and results}

\subsection{Observations and data analysis}

The VHE $\gamma$--ray observations were performed with the MAGIC 
telescopes located on the Canary Island of La Palma (28.8$^\circ$ N,
17.8$^\circ$ W at 2200 m.a.s.l). The two 17\,m telescopes use the 
imaging atmospheric Cherenkov technique, with a sensitivity of ($0.76\pm0.03)$\% of the Crab Nebula flux\footnote{In 50\,h of effective time in the medium energy range $>$ 290\,GeV, 
see Aleksi\'c et al. (2012c) for details.}. The energy threshold can be as low as 50\,GeV, a characteristic making the MAGIC telescopes well-suited for discovering and studying 
extragalactic VHE $\gamma$--ray sources. 

The BL Lac object 1ES 1727+502 was observed with the two MAGIC telescopes, using a hardware stereo trigger, between May 6th 
and June 10th 2011. During 14 nights 20.2 hours of data were collected. After a quality selection based on the event rate, excluding runs taken during adverse atmospheric 
conditions or with technical problems, the final data sample amounts to 14.0 hours. The effective time of this observation, corrected for 
the dead time of the trigger and readout systems is 12.6 hours. 
Parts of the data were taken under moderate moonlight and twilight conditions and were analysed together with the dark data (Britzger et al. 2009). The source was observed 
at zenith angles between 22$^\circ$ and 50$^\circ$.  

All the data were taken in the false-source tracking mode (wobble, Fomin et al. 1994), in which the 
telescopes were alternated every 20 minutes between two sky positions at 0.4$^\circ$ offset 
from the source.  

The data were analysed using the standard MAGIC analysis framework MARS as 
described in Moralejo et al. (2009) with additional adaptations incorporating the stereoscopic observations (Lombardi et al. 2011).  
The images were cleaned using timing information as described in Aliu et al. (2009) 
with absolute cleaning levels of 6 photoelectrons (so-called ``core pixels") and 3 photoelectrons
(``boundary pixels") for the first telescope and 9 photoelectrons and 4.5 photoelectrons for
 the second telescope. The images were parametrised in each telescope separately following the prescription of Hillas (1985).

We reconstructed the shower arrival direction with the
random forest regression method (RF DISP method, Aleksi\'c et al.
2010) which was extended using stereoscopic information such as the
height of the shower maximum and the impact distance of the shower on
the ground (Lombardi et al. 2011). 

For the gamma-hadron separation the random forest method was used (Albert et al. 2008b). 
In the stereoscopic analysis image parameters of both telescopes are used as well 
as the shower impact point and the shower height maximum. We additionally rejected events  
whose reconstructed source position in each telescope differs by 
more than 0.05 degree$^{2}$.
A detailed description of the stereoscopic MAGIC analysis can be found in
Aleksi\'c et al.\ (2012c).

\subsection{Results}

   \begin{figure}
   \includegraphics[width=9.3cm]{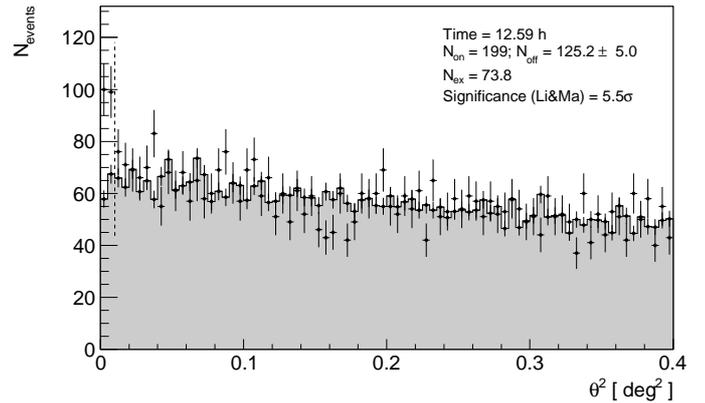}
 \caption{Distribution of the squared angular distance ($\theta^2$) between the source position and the reconstructed $\gamma$--ray direction for ON-source events (black points) and 
normalized OFF-source events (grey shaded area). The dashed line corresponds to the predefined region selected for the calculation of the significance of the detection. The respective 
statistics for ON and normalized OFF events are given in the figure.}
    \end{figure}

In the distribution of the squared angular distance between the catalogue position of 1ES 1727+502 and the reconstructed source position in the MAGIC data, the so-called $\theta ^2$ 
plot shown in Fig. 1, we find an excess ($N_{ex}$) of (73.8$\pm$15.0) events above the normalized background ($N_{off}$) of (125.2$\pm$5.0) events in the energy range above 150\,GeV. This corresponds to a 
significance of 5.5\,$\sigma$ calculated with formula 17 of Li \& Ma (1983), marking this observation as the first detection of 1ES 1727+502 in the VHE $\gamma$--ray regime. The 
integral flux above 150\,GeV is $(2.1\pm0.4)\%$ of the Crab Nebula flux. The fitted position of the excess is consistent with the catalogue coordinates 
(RA: 17.47184$^\circ$, Dec: 50.21956$^\circ$ as in Ma et al. 1998) within $ (0.032\pm0.015_{stat}\pm0.025_{sys}) ^\circ $, and thus compatible within the expected statistical 
and systematic errors (Aleksi\'c et al. 2012c). Comparing the extension of the excess to the point spread function of MAGIC ($\sim 0.1^\circ$, Aleksi\'c et al. 2012b), the source 
appears to be point-like. 

In order to take into account the effects of the finite energy resolution of the instrument, we unfolded the spectrum using the Forward Unfolding algorithm (described in Albert et al. 2007c). 
In the same procedure, the flux was corrected for the absorption due to the extragalactic background light pair-production using the model developed by 
Dom\'{i}nguez et al. (2011). The obtained differential flux  can be described by a power law function dF$/$dE $ = f_0(\mathrm{E}/300\,\mathrm{GeV})^{-\Gamma}$ with the 
following values of the parameters: flux normalization  $f_0 = (9.6\pm2.5)\times 10^{-12}\,\mathrm{cm}^{-2}\,\mathrm{s}^{-1}\,\mathrm{TeV}^{-1}$ and spectral
index $\Gamma = (2.7\pm0.5)$. We estimate a 10\% additional systematic uncertainty in the measured flux compared to Aleksi\'c et al. (2012c) due to the inclusion of 
moonlight and large zenith angle conditions in our data.

In Fig. 2 we present the VHE $\gamma$--ray light curve between 200\,GeV and 2\,TeV. In order to have a uniform distribution of days with observations in the bins 
and due to the weakness of the signal, a 14 day binning is applied starting from 2011 May 4. The resulting light curve has five observation nights in the first 
and last bin and four in the second bin. The emission is compatible with a constant flux of $(2.6\pm0.8)\times 10^{-12}\,\mathrm{cm}^{-2}\,\mathrm{s}^{-1}$. 
The relatively low probability of a constant flux (0.6\%, corresponding to a 2.5\,$\sigma$ rejection) might indicate variability below our detection threshold. The sparse binning and additional systematic errors due to moonlight and larger zenith angles can indeed fully explain this effect.

  \begin{figure}
   \includegraphics[trim=0 0 0 4.5mm, clip, width=9.0cm]{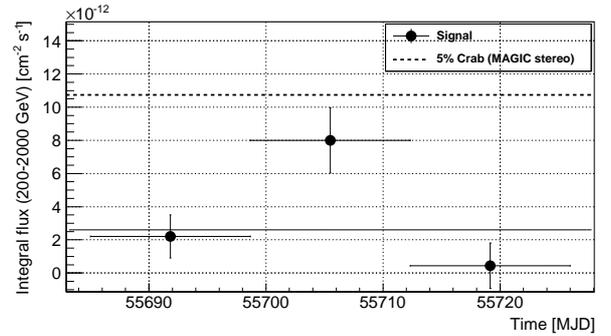}
 \caption{MAGIC light curve in the energy range from 200 GeV to 2 TeV. The Crab Nebula flux (Aleksi\'c et al. 2012c) scaled to 5\% is shown for comparison (dashed 
 line). The points correspond to the 14 days binned flux of 1ES 1727+502, and the error bars represent the statistical error only. The line represents the average flux during the entire observing period. The 
 probability of a constant flux is 0.6\% and the reduced $\chi^{2}$ with the number of degrees of freedom $n_{dof}$ of the fit assuming a constant flux is 10.12/2. }
    \end{figure} 

\section{Multiwavelength properties}

\subsection{Optical observations and results}

\begin{figure*}
\includegraphics[trim=1mm 8mm 0 0, clip,  width=6.5cm, angle=270]{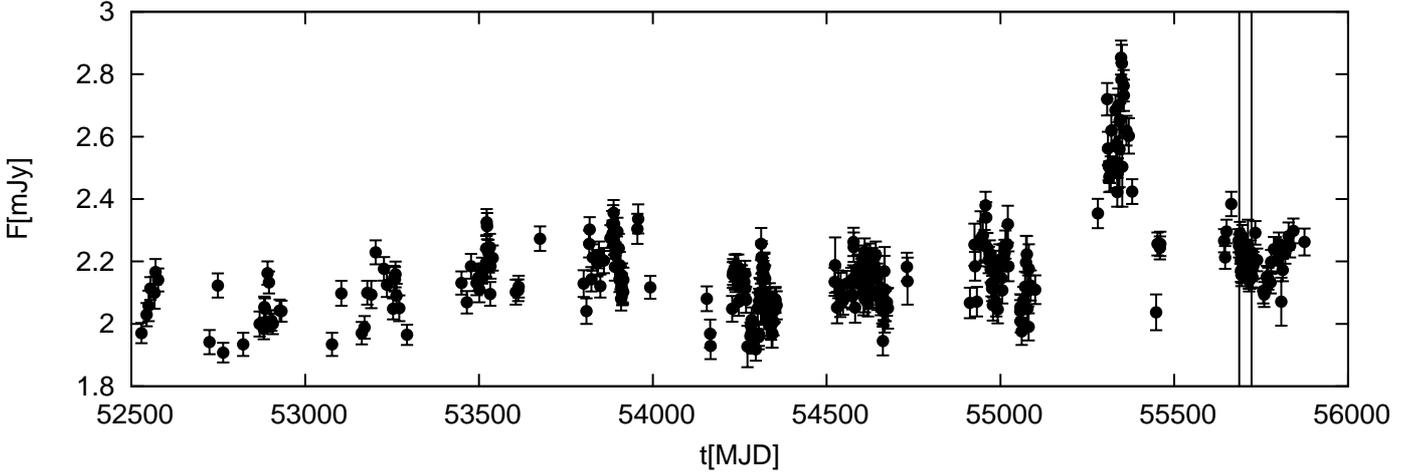}
\caption{10 years light curve in the optical \textit{R}-band from the Tuorla blazar monitoring program. The contribution of the host galaxy ($1.25\pm0.06\,$mJy) has not been subtracted. Vertical lines indicate beginning and end of the MAGIC observing window in 2011. See text for details.}
    \end{figure*}  
 
1ES 1727+502 has been observed continually in the optical \textit{R}-band as part of the Tuorla blazar monitoring program\footnote{http://users.utu.fi/kani/} for almost ten years, starting from 2002. The observations were carried out with the 1\,m Tuorla telescope and 35\,cm KVA telescope in La Palma. The brightness of the object was inferred from calibration stars in the same CCD-frames as 1ES 1727+502 using differential photometry and comparison star magnitudes from Fiorucci \& Tosti (1996). The magnitudes are converted to fluxes using the standard formula and values from Bessell (1979).

1ES 1727+502 has a bright host galaxy, contributing $>50\%$ to the flux in the optical \textit{R}-band (Nilsson et al. 2007). To derive the $\nu{\mathrm F}_\nu$ in the optical band, this contribution is subtracted from the measured flux and in addition the brightness was corrected for galactic absorption by \textit{R}=0.079\,mag (Schlegel et al. 1998). The average $\nu{\mathrm F}_\nu$ during the MAGIC observations  corresponds to $(4.93\pm0.2) \times 10^{-12}\,\mathrm{erg}\,\mathrm{cm}^{-2}\,\mathrm{s}^{-1}$.

Overall, the source showed mainly quiescent behaviour (as shown in Fig. 3) with the exception of an increased R-band flux starting in March-April 2010, with a peak value of $2.85\pm0.05$\,mJy on 2010 May 31, which exceeded the trigger criteria ($>$50\% above the long-term average) for MAGIC observations. However, as discussed in the introduction, the adverse atmospheric conditions forced us to discard the MAGIC data. The source had almost returned to its quiescent flux, $2.0 - 2.2$\,mJy, in September 2010 and remained in this state also during the MAGIC observations performed in 2011.

\subsection{Analysis and results of the \textit{Swift} archival data}

The \textit{Swift} Gamma-Ray Burst observatory, launched in November 2004 (Gehrels et al. 2004), is
equipped with three telescopes, the Burst Alert Telescope (BAT; Barthelmy et al. 2005), which covers the 15$-$150\,keV range, the X-ray
telescope (XRT; Burrows et al. 2005) covering the 0.3$-$10\,keV energy band, and the UV/Optical Telescope (UVOT; Roming
et al. 2005) covering the 1800$-$6000\,$\AA{}$ wavelength range. Unfortunately, there are no simultaneous \textit{Swift} observations during the MAGIC observing window. We have 
thus used archival data from April 5th and May 1st 2010. The data have been processed with standard procedures using the publicly available tools of the HEASoft package distributed by HEASARC. 

The results from \textit{Swift}/XRT are summarized in Table 1. The data have been fitted with a simple power law, in the range between 0.5$-$10\,keV. The flux is stable within $\sim 30\%$ during 
this period.     
\textit{Swift}/UVOT observations were performed during the same dates but only one of the observations, on April 5th, 2010 (MJD 55291.96182), contains all filters (\textit{V, B, U, W1, M2, W2}). We 
therefore used only this dataset for the compilation of the SED. The host galaxy contribution in the \textit{V} and \textit{B} bands was extrapolated from the \textit{R}-band values from 
Nilsson et al. (2007) using the galaxy colours at $z = 0$ from Fukugita et al. (1995). The the host galaxy contribution in the \textit{U} and ultraviolet bands is negligible. All the observed
magnitudes have been corrected for Galactic extinction E(\textit{B-V}) = 0.029\,mag (Schlegel et al. 1998), \textit{R} = 0.079, \textit{V} = 0.098, \textit{B} = 0.127, \textit{U} = 0.160 (taken 
from NED\footnote{http://ned.ipac.caltech.edu/}), and for the UV data \textit{W1} = 0.185, \textit{M2} = 0.272 and \textit{W2} = 0.243 using the curve from Fitzpatrick \& Massa (1999) and the 
central wavelengths from Poole et al. (2008). Final magnitudes have been converted into $\nu F_{\nu}$ and are summarized in Table 2.  

These archival \textit{Swift}/UVOT data were taken on 2010 April 5 when the optical flux was already increasing but before it reached the highest value, on 2010 May 31. 
Unfortunately there were no simultaneous observation with the KVA telescope but the \textit{R}-band 
SED point has a value of the flux, $4.93\pm0.2 \times 10^{-12}\,\mathrm{erg}\,\mathrm{cm}^{-2}\,\mathrm{s}^{-1}$, comparable to the spectral points obtained 
from \textit{Swift}/UVOT data (see Table 2). Consequently, the archival \textit{Swift}/UVOT can be regarded as representative of the baseline optical-UV flux and be included 
in the compilation of the multiwavelength SED.

\begin{table*}
\centering
\begin{tabular}{c c c c c}
  \hline \hline
  Observation date (MJD) & Observation time [ks] & Flux ($2-10\,$ keV) [$10^{-12}$ erg cm$^{-2}$ s$^{-1}$] & photon index & $\chi_{red} ^2(n_{dof})$ \\ \hline

55291.69584 & 2181.82 & 8.9$\pm$0.6 & 2.1$\pm$0.1 & 1.28(30) \\ 
55291.96182 & 1457.44 & 7.6$\pm$0.7 & 2.3$\pm$0.1 & 0.62(21) \\ 
55317.53682 & 1689.39 & 6.2$\pm$0.8 & 2.2$\pm$ 0.1 & 1.14(17) \\ 
\hline
\end{tabular}

\caption{Results of \textit{Swift}/XRT observations: observation date in MJD, exposure time, integral flux in the energy range $2-10\,$ keV, photon index of a simple power law fit function, reduced $\chi^{2}$ with the number of degrees of freedom $n_{dof}$.}

\end{table*}

\begin{table}
\centering
\begin{tabular}{c c}
  \hline \hline
  Band &  Flux [$10^{-12}$ erg cm$^{-2}$ s$^{-1}$]\\ \hline
\textit{V} & 5.4$\pm$ 0.7 \\ 
\textit{B} & 6.0$\pm$0.7 \\ 
\textit{U} & 6.7$\pm$0.4  \\ 
\textit{W1} &  6.0$\pm$0.3 \\ 
\textit{M2} &  6.3$\pm$0.3 \\ 
\textit{W2} &  7.2$\pm$0.3 \\ 
\hline

\end{tabular}

\caption{Results of \textit{Swift}/UVOT observations from 2010 April 5. }

\end{table}

\subsection{\emph{Fermi}-LAT data analysis and results}

1ES 1727+502 has been observed with the pair conversion Large Area Telescope (LAT) aboard \textit{Fermi} operating in the energy range from 20\,MeV up to  energies beyond 300\,GeV 
(Atwood et al. 2009, Abdo et al. 2012). In survey mode, the \textit{Fermi}-LAT scans the entire sky every three hours. 
The data sample used for this analysis covers observations from August 5th, 2008 to August 5th, 2011 and was analysed with the standard analysis tool {\it gtlike}, part  of the \textit{Fermi} ScienceTools software package (version 09-27-01) available from the \textit{Fermi} Science Support Center (FSSC). 
Only events belonging to the Pass7-V6 Source class and located within $10^{\circ}$ of 1ES 1727+502 were  selected. Moreover, to reduce the contamination from the Earth-limb $\gamma$ rays produced by cosmic rays interacting with the upper atmosphere, the data were restricted to a maximal zenith angle of $100^{\circ}$ and time periods when the spacecraft rocking angle exceeded $52^{\circ}$ were excluded.
To extract the source spectral information we used the standard background models publicly available at the FSSC website\footnote{http://fermi.gsfc.nasa.gov/ssc/data/access/lat/BackgroundModels.h\\tml}.
The background template separately models the Galactic diffuse emission and an isotropic diffuse emission, resulting from extragalactic isotropic emission and residual instrumental background. The normalization of these two templates were left free in the subsequent spectral fitting.
Sources from the 2FGL catalogue (Nolan et al. 2012) located within $15^{\circ}$ of 1ES 1727+502 were incorporated in the model of the region by setting their spectral models and the initial parameters for the modelling to those reported in the 2FGL catalogue. 
In the fitting procedure the parameters of sources located within $10^{\circ}$ radius centred on the source of interest were allowed to vary freely while parameters of sources located within the $10^{\circ}$-$15^{\circ}$ annulus were fixed. 
The model of the region around the source was forward folded with the post-launch instrument response functions P7SOURCE V6 and an unbinned maximum likelihood analysis was performed against the flight dataset between 300\,MeV -- 300\,GeV to derive the sources flux.
The uncertainties here reported in the LAT flux measurements are statistical only, systematic uncertainty in the LAT flux  can be derived from the systematic uncertainty on the effective area which is estimated to  be 10\% at 100\,MeV, 5\% at 560\,MeV and 10\% at 10\,GeV and above (Abdo et al. 2009).
\\
Since the source is not always significantly detected, flux upper limits at 95\% confidence level were calculated for each time bin where the test statistic (TS, it is 2 times the difference of the log(likelihood) with and without the source, see Mattox et al. 1996) value for the source was TS$<$4 or the number of predicted photons $N_{\mathrm{pred}}<3$. The light curve, from August 5th, 2008 to July 20th 2011, is 
presented in Fig. 4. Possible variations in the source emission have been tested following the same likelihood method described in the second \textit{Fermi} catalogue (Nolan et al. 2012). 
The result here obtained are consistent with a constant flux (TS$_{var}$=6 for 11 degrees of freedom), albeit a trend towards a higher flux in the 
last bin, partially coincident with the MAGIC observations, is evident. We also present in Fig. 5 the spectrum obtained from three months of observations centred 
around the MAGIC observing period. 
Compared to the average flux, in the energy range from 300\,MeV to 300\,GeV, measured in three years of observations ($3.5\pm0.5\times 10^{-9}\;$ph cm$^{-2}$s$^{-1}$), the flux measured in the three months around the MAGIC observations is higher ($7.2\pm1.9\times 10^{-9}\;$ph cm$^{-2}$s$^{-1}$), while the spectral indices are similar (1.90$\pm$0.08 and 2.0$\pm$0.2 respectively).
When performing the fit for the light curve and SED bins, the spectral indices of the sources were frozen to the 
best-fit values obtained from the time-independent analysis.

   \begin{figure}
   \includegraphics[width=9.0cm]{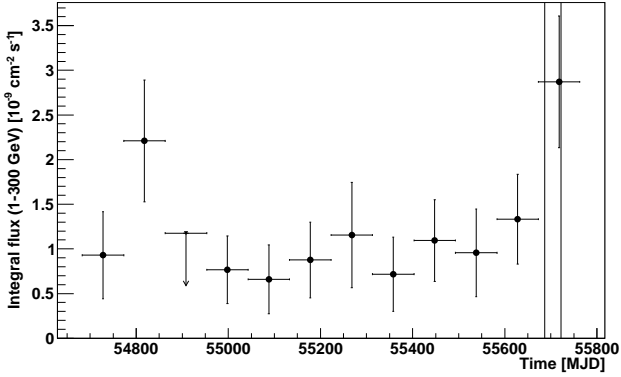}
\caption{Light curve with a binning of three months of the \textit{Fermi}--LAT data between 1\,GeV and 300\,GeV. The downward pointing arrows correspond to a 95$\%$ upper limit. The vertical lines indicate beginning and end of the MAGIC observing window in 2011. The emission is consistent with a constant flux, albeit a trend towards a higher flux in the last bin, partially coincident with the MAGIC observations, is evident.}
    \end{figure}

\subsection{Multiwavelength spectral energy distribution}

The quasi-simultaneous multiwavelength data described in the previous section have been used for the compilation of the SED, which has been modelled with a one-zone SSC model (Maraschi \& Tavecchio 2003). In this scenario, a blob of radius $R$ populated by relativistic 
electrons and filled with a tangled magnetic field of intensity $B$, is moving down the jet with a Doppler factor $\delta$. The electrons emit synchrotron radiation, producing the low-energy peak 
in the SED. The $\gamma$ rays are produced by the same electron population up-scattering the synchrotron photons, resulting in the second peak in the SED.

The electron spectrum is assumed to be described by $N(\gamma)=K \gamma^{-n_1}(1+\gamma/\gamma_b)^{n_1-n_2}$. The parameter values that give a good match between the SSC model and the SED data are: the Lorentz 
factors $\gamma_{min}=100, \gamma_{b}=3 \times 10^4, \gamma_{max}=6 \times 10^5$; the slopes $n_1=2, n_2=3.5$; and the electron density $K=8\times10^3\,$cm$^{-3}$.  The 
parameters that describe the astrophysical environment are the magnetic field $B=0.1\,$G, the radius $R=7\times 10^{15}\,$cm and the Doppler factor $\delta=15$ of the emitting region. 
These values are compatible with the values obtained with the sample analyzed in Tavecchio et al. 2010. 

\begin{figure}
   \includegraphics[trim=8.0mm 21mm 0 14.5mm, clip, width=9.3cm]{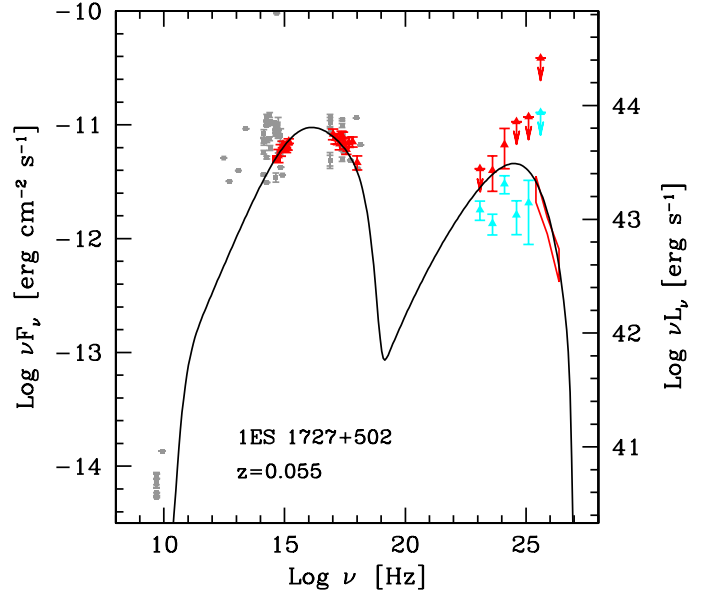}
 \caption{Multiwavelength spectral energy distribution fitted with a one zone synchrotron Self Compton model (Maraschi \& Tavecchio 2003). MAGIC observations, 
 (red butterfly) have been corrected for the extragalactic background light absorption using the model of Dom\'{i}nguez et al. (2011). The data used for the fit (red triangles) are: optical from KVA, archival UV and optical from \textit{Swift}/UVOT,  
 archival X-ray from \textit{Swift}/XRT, HE $\gamma$ rays from \textit{Fermi}-LAT (triangles, three months centred around the MAGIC observing period) and VHE $\gamma$ rays from MAGIC. We also show the 3 year LAT data (light blue triangles) and archival data (grey) from the ASI/ASDC archive ({\tt http://tools.asdc.asi.it/}).}
    \end{figure}

\section{Discussion}

The HBL 1ES 1727+502 shows little variability in the optical \textit{R}-band, is bright in the X-ray band, has a hard spectrum in the HE $\gamma$--ray band and,  
as shown in this paper, is visible in the VHE $\gamma$--ray range. The discovery of this source as VHE $\gamma$-ray emitter demonstrates the importance of combining 
data at different wavelengths, namely radio, optical, X-ray, and the recently opened {\it Fermi}--LAT energy range, to help identify potential VHE $\gamma$-ray emitters. The 
MAGIC detection indeed confirms the prediction made by Costamante \& Ghisellini (2002) and Donato et al. (2001) more than ten years ago, 
using X-ray, optical and radio data. Of the 33 sources in the list they compiled, 
21 have been already detected. They predicted a flux of $0.7\times 10^{-12}\,\mathrm{cm}^{-2}\,\mathrm{s}^{-1}$ above 300$\,$GeV and we observed a flux a factor 
of two higher ($1.6\times 10^{-12}\,\mathrm{cm}^{-2}\,\mathrm{s}^{-1}$).

Furthermore it is also interesting to compare this result with the excess seen in the stacked AGN sample observed by MAGIC in mono mode (Aleksi\'c et al. 2011). The 
spectral index measured for 1ES 1727+502 in the MAGIC energy range is compatible with the average spectral index of the stacked AGN sample: (2.7$\pm$0.5) compared 
to (3.2$\pm$0.5). Finally, when compared to the sample of all blazars detected in VHE $\gamma$-rays, its spectral index has
the value of a typical BL Lac, while the flux is one of the lower fluxes detected so far 
(Becerra et al. 2012; Becerra et al. 2013).  

We have interpreted the emission with a single-zone SSC model and find that the model parameters are compatible with those obtained for other sources of the HBL 
class. We investigated the multiwavelength variability of the source. During MAGIC observations the source was in a quiescent state in the 
optical band, and the {\it Fermi}--LAT data suggest (though not significantly) a flux enhancement during our observations compared to the three year averaged 
spectrum. 
We thus conclude that a study of the variability of this source, complemented with simultaneous multiwavelength observations, should be the focus of future 
observations. It will indeed help us in understanding not only the behaviour of this particular $\gamma$-ray emitter, but also the general features characterizing 
the HBL class of blazars.

\begin{acknowledgements}

 \small
We would like to thank the Instituto de Astrof\'{\i}sica de
Canarias for the excellent working conditions at the
Observatorio del Roque de los Muchachos in La Palma.
The support of the German BMBF and MPG, the Italian INFN, 
the Swiss National Fund SNF, and the Spanish MICINN is 
gratefully acknowledged. This work was also supported by the CPAN CSD2007-00042 and MultiDark
CSD2009-00064 projects of the Spanish Consolider-Ingenio 2010
programme, by grant DO02-353 of the Bulgarian NSF, by grant 127740 of 
the Academy of Finland,
by the DFG Cluster of Excellence ``Origin and Structure of the 
Universe'', by the DFG Collaborative Research Centers SFB823/C4 and SFB876/C3,
and by the Polish MNiSzW grant 745/N-HESS-MAGIC/2010/0.

The \textit{Fermi}-LAT Collaboration acknowledges support from a number of agencies and institutes for both development and the operation of the LAT as well as scientific data analysis. These include NASA and DOE in the United States, CEA/Irfu and IN2P3/CNRS in France, ASI and INFN in Italy, MEXT, KEK, and JAXA in Japan, and the K.~A.~Wallenberg Foundation, the Swedish Research Council and the National Space Board in Sweden. Additional support from INAF in Italy and CNES in France for science analysis during the operations phase is also gratefully acknowledged.

\end{acknowledgements}

\end{document}